\documentclass[3p,twocolumn]{elsarticle}

\usepackage{amssymb}
\usepackage{graphicx}
\usepackage{epsfig}
\usepackage{amsmath}
\usepackage{dcolumn}
\usepackage{bm}
\usepackage{dcolumn}
\usepackage{color}
\usepackage{verbatim}
\usepackage{paralist}
\usepackage{url}
\usepackage{lineno}

\journal{Physics Letters B}

\begin{document}

\begin{frontmatter}



\title{Study of solar and other unknown anti-neutrino fluxes with Borexino at LNGS}


\author[Milano]{G.~Bellini\fnref{label5}}
\fntext[label5]{spokesperson: Gianpaolo.Bellini@mi.infn.it}
\author[PrincetonChemEng]{J.~Benziger}
\author[Milano]{S.~Bonetti}
\author[Milano]{M.~Buizza~Avanzini}
\author[Milano]{B.~Caccianiga}
\author[UMass]{L.~Cadonati}
\author[Princeton]{F.~Calaprice}
\author[Genova]{C.~Carraro}
\author[Princeton]{A.~Chavarria}
\author[Moscow]{A.~Chepurnov}
\author[Milano]{D.~D'Angelo}
\author[Genova]{S.~Davini}
\author[Peters]{A.~Derbin}
\author[Kurchatov]{A.~Etenko}
\author[Dubna,LNGS]{K.~Fomenko}
\author[Milano]{D.~Franco}
\author[Princeton]{C.~Galbiati\fnref{label2}}
\fntext[label2]{Also at Fermi National Accelerator Laboratory, Batavia, IL 60510, USA}
\author[LNGS]{S.~Gazzana}
\author[LNGS]{C.~Ghiano}
\author[Milano]{M.~Giammarchi}
\author[Munich]{M.~Goeger-Neff}
\author[Princeton]{A.~Goretti}
\author[Genova]{E.~Guardincerri}
\author[Virginia]{S.~Hardy}
\author[LNGS]{Aldo~Ianni}
\author[Princeton]{Andrea~Ianni}
\author[Virginia]{M.~Joyce}
\author[Kiev]{V.V.~Kobychev}
\author[Dubna]{D.~Korablev}
\author[LNGS]{Y.~Koshio}
\author[LNGS]{G.~Korga}
\author[APC]{D.~Kryn}
\author[LNGS]{M.~Laubenstein}
\author[Princeton]{M.~Leung}
\author[Munich]{T.~Lewke}
\author[Kurchatov]{E.~Litvinovich}
\author[Princeton]{B.~Loer}
\author[Milano]{P.~Lombardi}
\author[Milano]{L.~Ludhova}
\author[Kurchatov]{I.~Machulin}
\author[Virginia]{S.~Manecki}
\author[Heidelberg]{W.~Maneschg}
\author[Genova]{G.~Manuzio}
\author[Munich]{Q.~Meindl}
\author[Milano]{E.~Meroni}
\author[Milano]{L.~Miramonti}
\author[Krakow]{M.~Misiaszek}
\author[LNGS,Princeton]{D.~Montanari}
\author[Peters]{V.~Muratova}
\author[Munich]{L.~Oberauer}
\author[APC]{M.~Obolensky}
\author[Perugia]{F.~Ortica}
\author[Genova]{M.~Pallavicini}
\author[LNGS]{L.~Papp}
\author[Milano]{L.~Perasso}
\author[Genova]{S.~Perasso}
\author[UMass]{A.~Pocar}
\author[Virginia]{R.S.~Raghavan}
\author[Milano]{G.~Ranucci}
\author[LNGS]{A.~Razeto}
\author[Milano]{A.~Re}
\author[Genova]{P.~Risso}
\author[Perugia]{A.~Romani}
\author[Virginia]{D.~Rountree}
\author[Kurchatov]{A.~Sabelnikov}
\author[Princeton]{R.~Saldanha}
\author[Genova]{C.~Salvo}
\author[Heidelberg]{S.~Sch\"onert}
\author[Heidelberg]{H.~Simgen}
\author[Kurchatov]{M.~Skorokhvatov}
\author[Dubna]{O.~Smirnov}
\author[Dubna]{A.~Sotnikov}
\author[Kurchatov]{S.~Sukhotin}
\author[LNGS]{Y.~Suvorov}
\author[LNGS]{R.~Tartaglia}
\author[Genova]{G.~Testera}
\author[APC]{D.~Vignaud}
\author[Virginia]{R.B.~Vogelaar}
\author[Munich]{F.~von~Feilitzsch}
\author[Munich]{J.~Winter}
\author[Krakow]{M.~Wojcik}
\author[Princeton]{A.~Wright}	
\author[Munich]{M.~Wurm}
\author[Princeton]{J.~Xu}
\author[Dubna]{O.~Zaimidoroga}
\author[Genova]{S.~Zavatarelli}
\author[Heidelberg]{G.~Zuzel}
\author{\\(Borexino Collaboration)}

\address[APC]{Laboratoire AstroParticule et Cosmologie, 75231 Paris cedex 13, France}
\address[Dubna]{Joint Institute for Nuclear Research, 141980 Dubna, Russia}
\address[Genova]{Dipartimento di Fisica, Universit\`a e INFN, Genova 16146, Italy}
\address[Heidelberg]{Max-Planck-Institut f\"ur Kernphysik, 69029 Heidelberg, Germany}
\address[Kiev]{Institute for Nuclear Research, 03680 Kiev, Ukraine}
\address[Krakow]{M.~Smoluchowski Institute of Physics, Jagiellonian University, 30059 Krakow, Poland}
\address[Kurchatov]{RRC Kurchatov Institute, 123182 Moscow, Russia}
\address[LNGS]{INFN Laboratori Nazionali del Gran Sasso, SS 17 bis Km 18+910, 67010 Assergi (AQ), Italy}
\address[Milano]{Dipartimento di Fisica, Universit\`a degli Studi e INFN, 20133 Milano, Italy}
\address[Moscow]{Institute of Nuclear Physics, Lomonosov Moscow State University, 119899, Moscow, Russia}
\address[Munich]{Physik Department, Technische Universit\"at Muenchen, 85747 Garching, Germany}
\address[Perugia]{Dipartimento di Chimica, Universit\`a e INFN, 06123 Perugia, Italy}
\address[Princeton]{Physics Department, Princeton University, Princeton, NJ 08544, USA}
\address[PrincetonChemEng]{Chemical Engineering Department, Princeton University, Princeton, NJ 08544, USA}
\address[Peters]{St. Petersburg Nuclear Physics Institute, 188350 Gatchina, Russia}
\address[UMass]{Physics Department, University of Massachusetts, Amherst, MA 01003, USA}
\address[Virginia]{Physics Department, Virginia Polytechnic Institute and State University, Blacksburg, VA 24061, USA}


\begin{abstract}
We report on the search for anti-neutrinos of yet unknown origin with
the Borexino detector at the Laboratori Nazionali del Gran Sasso.  In
particular, a hypothetical anti-neutrino flux from the Sun is
investigated.  Anti-neutrinos are detected through the
neutron inverse $\beta$ decay reaction in a large liquid organic scintillator
target.  We set a new upper limit for a hypothetical solar $\bar{\nu}_e$ flux of 760 ${\rm cm}^{-2}{\rm s}^{-1}$, obtained assuming an undistorted solar $^8$B energy spectrum.
This corresponds to a limit on the transition probability of solar neutrinos to
anti-neutrinos of $1.3\times10^{-4}$ (90\%
C.L.) for $E_{\bar{\nu}}>1.8$ MeV, covering the entire $^{8}$B spectrum.
Best differential limits on anti-neutrino
fluxes from unknown sources are also obtained between the detection
energy threshold of 1.8 MeV and 17.8 MeV with more than 2 years of data.
\end{abstract}


\end{frontmatter}

 \linenumbers



The Borexino Collaboration has recently published a measurement of
electron anti-neutrino fluxes at the Laboratori Nazionali del Gran
Sasso (LNGS)~\cite{BrxGeo}.  Contributions from two known sources were
observed: 1) $\bar{\nu}_e$'s produced in nuclear reactors
and 2) geo-neutrinos, produced in $\beta$ decays of isotopes along the
decay chains of long-lived $^{238}$U and $^{232}$Th distributed within
the Earth's interior. An $\bar{\nu}_e$ rate of 4.3$_{-1.4}^{+1.7}$
events/(100 ton yr) was measured from nuclear reactors, consistent
with an expected rate of 5.7$\pm0.3$ events/(100 ton yr).
Geo-neutrinos were identified at a rate of 3.9$_{-1.3}^{+1.6}$
events/(100 ton yr).

In this letter we present a study of other possible anti-neutrino sources.
These include the search for hypothetical solar anti-neutrinos and the investigation of other, unspecified and model-independent $\bar{\nu}_e$ fluxes.
 A weak anti-neutrino flux from the Sun arising from $\bar{\nu_{e}}\rightarrow\nu_{e}$ conversion cannot be completely excluded with current experimental data.
In particular, the interplay of flavor oscillations and spin flavor precession (SFP) induced by solar magnetic fields on Majorana neutrinos with
 sizable electric or magnetic transition moments \cite{Sch81}-\cite{Giu09} could lead to the appearance of
an $\bar{\nu}_e$ admixture in the solar neutrino flux.

The current best limit on the solar anti-neutrino flux is
$\phi_{\bar{\nu_{e}}}<370$ cm$^{-2}$s$^{-1}$ ($90\%$ C.L.), reported
by KamLAND \cite{KamLANDanu}.
The analysis was performed in the $8.3<E_{\bar{\nu_{e}}}<14.8\:$MeV
energy range. Assuming the undistorted solar  $^8$B spectrum, the limit on the anti-neutrino flux scaled to the entire energy range is $\phi_{\bar{\nu_{e}}}<1250$ cm$^{-2}$s$^{-1}$ ($90\%$ C.L.), and a limit on the conversion probability
$p_{\nu\rightarrow\bar{\nu}}<2.8\times10^{-4}$ ($90\%$ C.L.) was set using the $^{8}$B theoretical solar neutrino flux of
$5.05_{-0.16}^{+0.20}\times10^{6}$ cm$^{-2}$s$^{-1}$~\cite{bahcall} \footnote{In order to make the comparison of results clearer, we underline that the value for the $^8$B flux used in current work is $5.88\times10^{6}$ ${\rm {\rm cm}}^{-2}{\rm {\rm {\rm s}}}^{-1}$~\cite{serenelli}.}.
The expected background from known sources for this search in the energy range of
interest was $1.1\pm0.4$ events/(0.28 kt yr) and no candidate events were
observed in 185 days of data taking.
%

Although a smaller detector than KamLAND, Borexino has a competitive sensitivity
in the high energy portion of the anti-neutrino energy spectrum,
above reactor anti-neutrinos.  At lower energies,  Borexino compensates its size disadvantage with a significantly lower anti-neutrino background,
due to its large distance from nuclear power plants (the average baseline is $\sim1000$ km) and with lower intrinsic radioactive background.

Borexino~\cite{sctech}-\cite{Borexino2} is a large volume, unsegmented organic liquid scintillator detector located underground
at the Laboratori Nazionali del Gran Sasso, Italy. Primarily designed for a real-time, high precision measurement of the
mono-energetic (862 keV) $^7$Be solar neutrino flux via $\nu-e$ elastic scattering interactions in the scintillator, Borexino is
also an extremely sensitive anti-neutrino detector. It began data taking in May 2007 and has already presented a measurement of
the $^7$Be solar neutrino flux with 10\% precision \cite{bx2}. Because both $\beta$ and $\gamma$ interactions are essentially
indistinguishable from the sought-after neutrino induced events, the measurement was possible thanks to the high purity from
radioactive contaminants achieved in the scintillator and surrounding detector components, (1.6$\pm$0.1)$\times10^{-17}$~g/g for
$^{238}$U and (6.8$\pm$1.3)$\times10^{-18}$~g/g for $^{232}$Th~\cite{bx2}.

A central spherical core of 278~tons (design value) of organic liquid scintillator (LS) solution, constituted of pseudocumene
solvent (1,2,4-trimethylbenzene C$_{6}$H$_{3}$(CH$_{3}$)$_{3}$) doped with PPO fluor (2,5- diphenyloxazole, C$_{15}$H$_{11}$NO)
with a concentration of 1.5 g/l, is contained within a 8.5 m-diameter thin transparent nylon vessel (Inner Vessel, IV) and
viewed by 2212 large area 8" photomultiplier tubes (PMTs) defining the inner detector (ID) and providing 34\% geometric
coverage.  The scintillator is immersed in $\sim$1000 tons of pseudocumene buffer fluid, divided into two regions by a second
transparent nylon vessel 11.5 m in diameter, which prevents radon gas to permeate into the scintillator.  Scintillator and buffer
fluid are contained within a 13.7 m-diameter stainless steel sphere (SSS) on which the inward-looking PMTs are mounted.  DMP
(dimethylphtalate) dissolved in the buffer fluid quenches undesired scintillation from residual radioactivity contained in the
SSS and PMTs.  The SSS is immersed in a large water tank (WT) for further shielding from high energy $\gamma$ rays and neutrons
emerging from the surrounding rock.  The WT is instrumented as a \v{C}erenkov muon detector (outer detector, OD) with additional
208 PMTs, particularly important for detecting muons skimming the central detector inducing signals in the energy region of
interest for neutrino physics.  A detailed description of the Borexino detector can be found in Refs.~\cite{Borexino,Borexino2}.

A high light yield of $\simeq500$ photoelectrons~(p.e.) detected for
every 1 MeV of electron energy deposited gives an energy resolution of
$\sim$5\% at 1~MeV.  The position of each scintillation event in
Borexino is determined from the timing pattern of hit PMTs with a
spatial reconstruction algorithm. The code has been tuned using data
from several calibration campaigns with radioactive sources inserted
at different positions inside the detector. A maximum deviation of 5 cm
between the reference and reconstructed source positions is
observed at a radius of $\sim4$ m, close to the bottom of the IV.

In Borexino, anti-neutrinos are detected via the neutron inverse
$\beta$ decay reaction, $\bar{\nu_{e}}+p\rightarrow n+e^{+}$, with a
kinematic threshold $E_{\bar{\nu}}>1.8$ MeV.
The cross section for this process is much higher than for
$\bar{\nu}-e$ elastic scattering, making it the dominant anti-neutrino
interaction in proton-rich water-\v{C}erenkov and liquid scintillator
detectors.
This process also offers an experimentally unique signature given by
the close time sequence of correlated events.  The positron promptly
annihilates emitting two 511 keV $\gamma$-rays, providing the prompt event, with
a visible energy of $E_{\rm prompt} = E_{\bar{\nu}}-0.782$ MeV (the
scintillation light related to the proton recoil is highly quenched
and negligible).  The neutron quickly thermalizes and is then captured
by a proton after a time characteristic of the medium, via the
reaction $n+p\rightarrow d+\gamma$.  For protons, the de-excitation
$\gamma$ ray is 2.2 MeV and constitutes the delayed event.  For the
Borexino scintillator the mean capture time was measured to be
$\sim256$ $\mu s$.  The coincident nature of anti-neutrino events
allows for the detection of relatively few events with high
significance.  Incidentally, the 2.2 MeV photon is detected
with low efficiency in water-\v{C}erenkov detectors.

For the present analysis we used two antineutrino candidates selection criteria.
With data set A, we selected anti-neutrinos candidates in the entire scintillator volume from the data collected between May 2007 and June 2010.
The live time for set A after all analysis cuts is 736 days.
Data set B coincides with the one used for the geo-neutrino analysis \cite{BrxGeo} and includes data taken between December 2007 and December 2009, for a total 482
days of live time.
Analogously to what was done for the geo-neutrino analysis,
a fiducial volume cut, detailed below, was introduced for data set B in order to suppress neutron background from $^{13}{\rm C}(\alpha, n)^{16}{\rm O}$ reactions initiated by $^{210}$Po $\alpha$ decays in the non-scintillating buffer fluid surrounding the scintillator.

The following anti-neutrino candidates selection criteria have been
defined based on calibration data and Monte Carlo (MC) simulations and
applied to the data.  All cuts apply to both data sets unless
otherwise noted.
\begin{enumerate}
\item $Q_{\rm prompt}>410$ p.e.; for reference, $Q=438\pm2$
p.e. corresponds to the positron annihilation at rest with $E=1.022$
MeV released at the center of the detector.
\item $700<Q_{\rm delayed}<1250$ p.e.; 2.2 MeV $\gamma$'s deposit
$1060\pm5$ p.e.  at the detector's center; the lower limit is
justified because photons at the edge of the scintillator can escape
depositing none or only a fraction of their total energy).
%
%
\item $20<\Delta T<1280\mbox{ }$$\mu$s, where $\Delta$T is the time
between prompt and delayed event. The upper limit is 5 times the
mean neutron capture time and guarantees good acceptance.  The lower
limit excludes double cluster events, {\sl i.e.} events that fall
within the same data acquisition gate, which present higher background
contamination.
\item Reconstructed distance between prompt and delayed events:
$\Delta R<1$ m for data set B, $\Delta R<1.5$ m for data set A to
increase the acceptance.
\item $R_{\rm prompt}<R_{\rm IV}(\theta,\phi)-0.25$ m for data
set B, where $R_{\rm prompt}$ is the reconstructed radius for the prompt event
and $R_{\rm IV}(\theta,\phi)$ is the inner vessel radial size in the
direction ($\theta,\phi$) of the event (the IV is not exactly spherical;
its true shape is reconstructed using seven CCD cameras mounted on the SSS).
No volume cut is applied for data set A.
\item All tagged muon events are dropped.  Muons can cause events which mimick
$\bar{\nu}_e$ events as illustrated in \cite{BrxGeo}.  Cosmic muons are
typically identified by the OD, but can also be distinguished from
point-like scintillation events by the pulse shape analysis of the ID
signal. The probability to miss a muon after identification by the OD and pulse shape analysis is $3\times10^{-5}$~\cite{BrxGeo}.
\item A 2 ms veto is applied after each muon which crosses the outer
but not the inner detector in order to suppress background from fast
neutrons produced along their tracks in water.
\item A 2 second veto is applied after each muon crossing the inner detector to
suppress the $\beta-n$ decaying cosmogenic isotopes $^{9}$Li
($\tau=260$ ms) and $^{8}$He ($\tau=173$ ms).
\end{enumerate}
Approximately 4300 muons traverse the ID every day, and an equivalent
amount cross the OD alone.  Cuts 6 and 7 thus introduce a modest dead
time.  Cut 8 corresponds to $\sim$8600 seconds of dead time
per day, or about 10\%.  A careful study of cuts 6-8 yielded a total
dead time of 10.5\%.  The combined acceptance $\epsilon$ of all other
cuts (1-5) is estimated by MC means at $(85\pm1)$\% for data
set B and $(83\pm1)$\% for data set A, mostly attributable to the
$\Delta T$ ($\epsilon=91.8$\%) and $\Delta R$ ($\epsilon\sim95$\%)
cuts, and partially due to the loss of $\gamma$ rays close to the IV
surface.

The energy of each event is reconstructed using the total amount of light registered by all PMTs of the detector (measured in photoelectrons, p.e.) and 
corrected with a position-dependent light collection function $f(x,y,z)$, which relates the light yield at point $(x,y,z)$ to that of the same event at the center of the detector, $Q(x,y,z)=f(x,y,z)\;Q_{0}$. An extensive calibration campaign with a set of gamma sources  has been performed (the energy of the gamma rays ranged between a few hundred keV to 9 MeV).  For events at the center of the detector, the total amount of light collected is linear with energy above 1 MeV~\cite{Pauli}. The $f(x,y,z)$ function was constructed using calibration data from sources deployed at different positions inside the detector and checked against MC simulations. The correction is up to 15\% at the poles of the detector, where scintillation light is shadowed the most by the conduits used to fill the detector. The estimated (systematic) uncertainly of the energy reconstruction procedure is about 2\%. It is mainly due to the uncertainty in the $f(x,y,z)$ function, with a small contribution from the stability of the energy scale of the detector. 
The most energetic anti-neutrino candidate event in the entire data set has a total light yield $Q_{\rm prompt}=2,996$ p.e. and is located within the fiducial volume defined for data set B.  $Q_{\rm prompt}=2996$ p.e. corresponds to $Q_{0}=2991$ p.e. at the center of the detector, or $E=6.22\pm0.12 \;(syst)$ MeV.  
We looked for an admixture of anti-neutrinos within the solar neutrino flux,
considering the case of energy independent conversion. Data Set A was used for this
analysis, with a threshold of 6.5~MeV for the visible energy (corresponding to an anti-neutrino energy of 7.3~MeV) . With this threshold, the expected background in Data Set A is mainly due to the high energy tail of reactor anti-neutrinos, is estimated at $N_{R}=0.31\pm0.02$ events. Additional backgrounds are hard to define and quantify, but in a conservative approach we can simply use the lowest possible value, i.e. zero background.
The absence of anti-neutrino candidates in the entire $270\pm3$ tons of scintillator
(as calculated by reconstructing the exact shape of the IV by means
of images taken with seven inward-looking CCD cameras mounted on the
SSS) during an observation time of $T=736$ days is then used to set
the following limit on a hypothetical solar anti-neutrino flux:

\begin{equation}
\phi_{\rm lim}=\frac{S_{\rm lim}}{\bar{\sigma}\cdot T\cdot N_{p}\cdot\epsilon}\;{\rm {\rm cm}}^{-2}{\rm {\rm {\rm s}}}^{-1},\label{eq:PhiLim}\end{equation}
where $N_{p}=(1.62\pm0.02)\times 10^{31}$ is the number of target
protons, $\bar{\sigma}$ is the average
cross section for neutron inverse beta decay~\cite{Strumia} weighted over the
$^8$B solar neutrino spectrum~\cite{b8Bahcall} in the energy range of interest, $S_{\rm
lim}$ is the maximum allowed signal at 90\% C.L. and $\epsilon =
0.83\pm0.01$ is the efficiency (constant within the chosen interval)
of inverse beta-decay detection for the chosen set of selected cuts, defined with MC
simulation.

Assuming an undistorted solar $^{8}$B neutrino energy spectrum $\bar{\sigma}(\rm{E}_{\bar{\nu}}>7.3)=6.0\times10^{-42}$ cm$^{2}$ 
and with $S_{\rm lim}=2.13$, obtained applying the Feldman-Cousins procedure
\cite{Stat} for the case of no observed events with 0.31 expected background events
we obtain:

\[
\phi_{\bar{\nu}}(^{8}{\rm B, E}> 7.3 {\rm MeV})<415\;{\rm {\rm cm}}^{-2}{\rm {\rm {\rm s}}}^{-1}\;(90\%\;{\rm C.L.}).\]
Part of the spectrum above 7.3~MeV corresponds to 42\% of the total
$^{8}$B neutrino spectrum~\cite{b8Bahcall}, equivalently $\phi_{\bar{\nu}}(^8{\rm
{B}})<990\;{\rm cm}^{-2}{\rm s}^{-1}$ at 90\% C.L. over the entire
$^8$B neutrino spectrum.
This corresponds to an average transition probability in this energy
range of $p_{\nu\rightarrow\bar{\nu}}<1.7\times10^{-4}$ obtained
assuming $\phi_{SSM}(^{8}{\rm {B}})=5.88\times10^{6}$ ${\rm {\rm
cm}}^{-2}{\rm {\rm {\rm s}}}^{-1}$~\cite{serenelli} (we do not take
into account the errors on the theoretical predictions of the $^{8}$B
neutrino flux). 

A limit was alternatively obtained using data set B over the
entire $^{8}$B spectrum.  The reduced statistics due to the lower target fiducial volume
($N_{p}=1.34\times10^{31}$), shorter data taking period ($T=482$ days)
and lower acceptance on $\Delta R$ with respect to data set A is partially compensated by the larger energy range and a
higher detection efficiency, $\epsilon=0.85\pm0.01$. Applying the fit
procedure developed for the geo-neutrino studies \cite{BrxGeo} and
following the $\chi^{2}$ profile as a function of the amount of
additional, hypothetical, anti-neutrinos under the assumption of an energy-independent
$^{8}$B solar neutrino conversion to anti-neutrino, we obtain $S_{\rm
lim}=2.90$ at 90\% C.L. From (\ref{eq:PhiLim}) and with $\bar{\sigma}(E_{\bar{\nu}}>1.8)=3.4\times10^{-42}\;{\rm cm}^2$:

\[
\phi_{\bar{\nu}}(^{8}{\rm B})<1820\;{\rm {\rm cm}}^{-2}{\rm {\rm {\rm s}}}^{-1}\;(90\%\;{\rm C.L.})\]
or $p_{\nu\rightarrow\bar{\nu}}<3.1\times10^{-4}$ in the energy range
above 1.8~MeV. This limit is weaker than that obtained by simple
scaling of the result from the study above 7.3~MeV.

\begin{figure}[t]
\begin{centering}
\includegraphics[width=0.4\paperwidth,height=0.35\paperwidth]{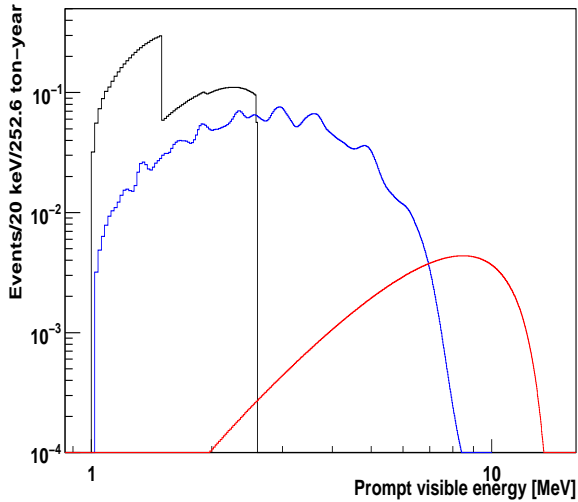}
\par\end{centering}

\caption{\label{AnuSpectrum} Energy spectra for electron
anti-neutrinos in Borexino. The horizontal axis shows the kinetic plus
the annihilation 1.022 MeV energy of the prompt positron event. Shown
are expected shapes for geo- (black line) and reactor (blue line)
anti-neutrinos normalized to the Borexino measured
values~~\cite{BrxGeo} for a 252.6 ton year exposure. Effects of
neutrino oscillations are included.  The spectral shape for
hypothetical $^{8}$B solar anti-neutrinos is shown in red and
normalized to our upper limit of
$1.3\times10^{-4}$$\phi_{SSM}(^{8}{\rm {B}})$ (see text for details)
}
\end{figure}

The strongest limit was obtained combining both data sets (below/above
7.3~MeV we used the data sets B/A, respectively).
We applied the fit procedure developed for the geo-neutrino studies
\cite{BrxGeo} with the $^{8}$B neutrino spectrum correctly weighted
for effective exposures below and above 7.3~MeV.
Following the $\chi^{2}$ profile with respect to the amount of additional hypothetical anti-neutrinos assuming an energy-independent
$^{8}$B neutrino conversion to anti-neutrino, we obtain $S_{\rm
lim}=2.07$ at 90\% C.L. (the corresponding fraction above 7.3~MeV is 1.64).
Using (\ref{eq:PhiLim}) with
$\bar{\sigma}=6.0\times10^{-42}\;{\rm cm}^2$
\[
\phi_{\bar{\nu}}(^{8}{\rm B, E}> 7.3 {\rm MeV})< 320\;{\rm {\rm cm}}^{-2}{\rm {\rm {\rm s}}}^{-1}\;(90\%\;{\rm C.L.,})\]
which in the total energy range corresponds to:
\[
\phi_{\bar{\nu}}(^{8}{\rm B})< 760\;{\rm {\rm cm}}^{-2}{\rm {\rm {\rm s}}}^{-1}\;90\%\;{\rm C.L.,}\]
or $p_{\nu\rightarrow\bar{\nu}}<1.3\times10^{-4}$ for the total $^{8}$B flux.
This limit is stronger than that obtained by simple scaling of the
result from the study above 7.3~MeV energy range.
Our limits for $\bar{\nu}$ flux with undistorted solar $^8$B neutrino spectrum are
summarized in Tab.~\ref{Tab:comparison}, where they are compared with
those reported by SuperKamiokaNDE~\cite{SuperK},
KamLAND~\cite{KamLANDanu} and SNO~\cite{SNO}. The upper limit on a hypothetical solar anti-neutrino
rate is illustrated against the Borexino measured
reactor and geo-neutrino rates~\cite{BrxGeo} in Fig.~\ref{AnuSpectrum}.

The limit on solar anti-neutrinos allows us to set limits on the
neutrino magnetic moment $\mu_{\nu}$ and on the strength and shape of
the solar magnetic field.  Assuming the SFP mechanism coupled with the
MSW-LMA solar neutrino solution, limits on the neutrino magnetic
moment $\mu_{\nu}$ can be obtained using the limits on the conversion
probability $p_{\nu \rightarrow \bar{\nu}}$, as shown
in~\cite{Akh03,Cha03,Bal05}. In general, the limit on $\mu_{\nu}$ depends on the unknown strength of the solar magnetic field $B$ in
the neutrino production region, and can be written as~\cite{Akh03}:

\[
\mu_{\bar{\nu}}\leq7.4\times10^{-7}\left(\frac{p_{\nu\rightarrow{\bar{\nu}}}}{{\rm sin}^{2}2\theta_{12}}\right)^{1/2}\frac{\mu_{B}}{B_{\perp}[{\rm kG}]}
\]
where $\mu_{B}$ is Bohr's magneton and $B_{\perp}$ is the transverse component of the
solar magnetic field at a radius $0.05R_{\odot}$ corresponding to the maximum of $^{8}$B-neutrino production. Using our experimental
limit $p_{\nu \rightarrow \bar{\nu}} = 1.3 \times 10^{-4}$ and ${\rm sin}^{2}2\theta_{12} = 0.86$~\cite{Fog08,Gon10} one obtains
 $\mu_{\nu} \le 9\times 10^{-9} B_{\perp}\; \mu_B$ (90\% C.L.).  Solar physics provides very limited knowledge on magnitude and shape
  of solar magnetic fields.  In accordance with~\cite{Kit08,Fri04,Ant08} the magnetic field in the core  can vary
  between  600 G and 7 MG. The higher value limits the magnetic moment to $\mu_{\nu} \le 1.4 \times 10^{-12} \mu_B$.

The spin flip could also occur in the convective zone of the Sun in which neutrinos traverse a region of random turbulent magnetic fields
\cite{Mir04,Mir04A,Fri05}. Using the expression for the conversion probability given in~\cite{Fri05,RR} and the solar neutrino mixing
parameters ${\rm cos}^2\theta_{12} = 0.688$ and $\Delta m^2 = 7.64 \times 10^{-5} {\rm eV}^2$~\cite{Fog08,Gon10}, the limit on the magnetic
moment can be written as:
\[
\mu_{\bar{\nu}}\leq8.2\times10^{-8}\;p_{\nu\rightarrow{\bar{\nu}}}^{1/2} \;B ^{-1}[{\rm kG}] \;\mu_{B}
\]
where B is the average strength of the turbulent magnetic field. Using conservative values for B of 10-20 kG~\cite{Mir04A,Fri05,RR,Fri03},
we obtain less stringent limits on the magnetic moment than current laboratory bounds.

Currently, the best limit on the neutrino magnetic moment, $\mu_{\nu}<3\times10^{-12}$ $\mu_{B}$, is obtained by imposing astrophysical
constraints that avoid excessive energy losses by globular-cluster stars \cite{Astro}.  The best direct limit is obtained with reactor
neutrinos, $\mu_{\bar{\nu}_e}<3.2\times10^{-11}$ $\mu_{B}$ \cite{Gemma} or $\mu_{\bar{\nu}_e}<5\times10^{-12}$ $\mu_{B}$ when atomic
ionization is taken into account \cite{AtI}. The best limit on the effective magnetic moment of solar neutrinos is close, 
$\mu_{\nu}<5.4\times10^{-11}$ (90\% C.L.) \cite{bx2}.  It was obtained by the Borexino Collaboration by studying the shape of the electron
recoil energy spectrum following elastic scattering from mono-energetic $^7$Be solar neutrinos.  These experimental limits on the neutrino
magnetic moments, together with reasonable assumptions on the distribution of turbulent magnetic fields in the Sun, corresponds to a
conversion probability $p_{\nu\rightarrow{\bar{\nu}}}\sim 10^{-6}$, about two orders of magnitude lower than the sensitivity of present
experiments.

The case of an undistorted $^8$B antineutrino spectrum is a special case of $\nu\rightarrow{\bar{\nu}}$ conversion.
For the more general case, a model-independent search for unknown anti-neutrino fluxes was
performed in 1 MeV energy bins for 1.8 MeV $< E_{\bar{\nu}}<17.8$ MeV.
This study was made possible by the extremely low background achieved
by Borexino. The analysis consisted in setting the limits on any contribution of unknown origin in the antineutrino spectrum registered by Borexino.
%
%
Data set B was used below 7.8 MeV (same region used in
our recent study of geo-neutrinos), where reactor and geo-neutrinos
positively contribute to the signal, with 21 detected candidate events
within 225.1 tons of scintillator during 482 days of data taking and
an average detection efficiency $\epsilon=0.85$ \cite{BrxGeo}.  Data
set A was used above 7.8 MeV where, as mentioned earlier, no events
were recorded allowing us to use the full scintillator volume and a
more inclusive $\Delta R$ value between the prompt and delayed events.
The live time was 736 days with an average detection efficiency
$\epsilon=0.83$.
Decrease in the detection efficiency at high energies due to muon
software tagging over--efficiency via ID pulse shape analysis was
considered and evaluated by MC simulations.
The robustness of the event binning was tested against the precision
of the light yield-to-energy conversion, which carries a 2\%
systematic uncertainty as mentioned earlier.  The number of events
assigned to each bin is not sensitive to the variation of parameters
(within 90\% C.L.) with the exception of bins 2 and 3.  One event
happens to be on the boundary of these two bins (within $\simeq$0.5\% energy interval) and we conservatively
assigned it to both in setting our limits.

In order to have conservative limits, the minimal expected number of
events in every bin has been calculated separately for reactor and
geo-neutrinos.
For geo-neutrinos we considered the  Minimal Radiogenic Earth
model~\cite{Mantovani2004}, which only includes the radioactivity from U and Th
in the Earth crust which can be directly measured in
rock-samples. More details on the calculation can be found in Ref.~\cite{BrxGeo}.

\begin{figure}
\begin{centering}
\includegraphics[width=0.5\textwidth,height=0.4\textheight]{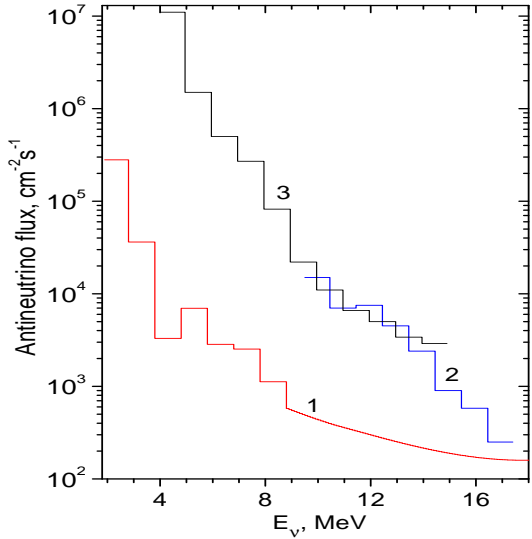}
\par\end{centering}
\caption{\label{AnuLimits} Upper limits on the monochromatic
$\bar{\nu}_e$ fluxes for: 1- the present Borexino data (red
line); 2- SuperKamiokaNDE (blue line) \cite{SuperK} and 3- SNO (black
line) \cite{SNO}. The regions above the lines are excluded.  The x-axis
is anti-neutrino energy.}
\end{figure}

The 90\% C.L. upper limits $S_{\rm lim}$ in the Feldman-Cousins approach for the first
eight bins are \{11.5, 6.48, 1.32, 4.93, 3.12, 3.96, 2.38, 2.43\}, respectively.
%
Above bin number 7 ($E>7.8$ MeV) S$_{\rm lim}=2.44$, obtained with the
Feldman-Cousins recipe for no observed events with zero background.
Model independent limits on anti-neutrino fluxes are illustrated in
Fig.~\ref{AnuLimits} and compared with SuperKamiokaNDE~\cite{SuperK}
and SNO~\cite{SNO} data.

\begin{table}[t!]
\caption{Limits on the $\bar{\nu}$ flux with undistorted $^8$B
spectrum for the Borexino, its prototype CTF, KamLAND, SNO and SuperKamiokaNDE
experiments.  Upper limits are given at 90\%~C.L. (see text for
details).}
%
\begin{center}
\begin{tabular}{lll}
\hline\hline
Experiment              & Measurement   &  Total $\phi_{\bar{\nu_{e}}}(^{8}$B) \\
                                                & threshold         & $90\%$ C.L.  \\
                        &[MeV]          & [cm$^{-2}$s$^{-1}$] \\
\hline
CTF~\cite{CTFanu}                &$>1.8$         &$<1.1\times10^{5}$ \\
SNO~\cite{SNO}              &$>4$       &$<4.09\times10^{4}$ \\
SuperK~\cite{SuperK}    &$>8$       & $<4.04\times10^{4}$ \\
KamLAND~\cite{KamLANDanu}   &$>8.3$         &$<1250$ \\
Borexino (this work)        &$>7.3$         &$<990$ \\
Borexino (this work)        &$>1.8$         &$<760$ \\
\hline\hline
\end{tabular}
\label{Tab:comparison}
\end{center}
\end{table}

As far as concerned the possible conversion of the low energy neutrino below
the 1.8 MeV inverse beta decay reaction threshold, including those that could originate
by SFP conversion of the abundant $^{7}$Be monoenergetic solar
neutrinos, the only available detection channel is the elastic scattering on electrons.
The recoil spectra for electrons elastically scattering off neutrino and anti-neutrinos are distinct, and we exploited such difference to search for an anti-neutrino admixture in the $^7$Be solar neutrino flux.  Possible deviations from the pure
$\nu-e$ electron recoil shape
due to electromagnetic interactions
were studied in Borexino previously
published data \cite{bx2}.  Following the changes in the $\chi^{2}$ profile with
respect to the addition of an anti-neutrino component we set a limit on
the conversion probability for $^7$Be solar neutrinos of
$p_{\nu_e\rightarrow\bar{\nu_{e}}}<0.35$ at 90\% C.L. The relatively
low sensitivity is in large part due to the strong anticorrelation
between the $\bar{\nu}-e$ elastic scattering spectrum and that of
$^{85}$Kr (a $\beta$ emitter which represents a significant residual background in Borexino) both left free in the analysis.  It is
likely that this limit could be improved following a purification
campaign of the scintillator.

In conclusion, Borexino has shown excellent sensitivity to
naturally-produced anti-neutrinos over a broad range of energies,
thanks to its unprecedented radiological purity and its location far
away from nuclear reactors.
New limits have been set on the possible $\bar{\nu}$ admixture in the solar
neutrino flux.
In particular,
$p_{\nu\rightarrow\bar{\nu}}<1.7\times10^{-4}$ (90\% C.L.) for
E$_{\bar{\nu}}>$7.3 MeV,
$p_{\nu\rightarrow\bar{\nu}}<1.3\times10^{-4}$ (90\% C.L.) for
the whole $^{8}$B energy region, and
$p_{\nu_{e}\rightarrow\bar{\nu_{e}}}<0.35$ (90\% C.L.) for 862 keV
$^{7}$Be neutrinos.
The best differential limits on anti-neutrino fluxes of unknown
origin between 1.8 and 17.8 MeV have also been set.

\section{Acknowledgements}

This work was funded by INFN (Italy), NSF (US Grant NSFPHY-0802646), BMBF (Germany), DFG (Germany, Grant OB160/1-1 and Cluster of Excellence “Origin and Structure of the Universe”), MPG (Germany), Rosnauka (Russia, RFBR Grant 09-02-92430), and MNiSW (Poland). This work was partially supported by PRIN 2007 protocol 2007 JR4STW from MIUR (Italy). O.Smirnov, L.Ludhova and A. Derbin aknowledge the support of Fondazione Cariplo.



\bibliographystyle{elsarticle-num}
\bibliography{<your-bib-database>}



\end{document}